\documentclass[twocolumn,floatfix,showpacs,superscriptaddress,prb,eqsecnum]{revtex4}
\usepackage{graphicx}
\usepackage{amsmath}
\usepackage{bm}
\begin{document}

\title{Effective mass and tricritical point for lattice fermions localized by a random mass}
\author{M. V. Medvedyeva}
\affiliation{Instituut-Lorentz, Universiteit Leiden, P.O. Box 9506, 2300 RA Leiden, The Netherlands}
\author{J. Tworzyd{\l}o}
\affiliation{Institute of Theoretical Physics, University of Warsaw, Ho\.{z}a 69, 00--681 Warsaw, Poland}
\author{C. W. J. Beenakker}
\affiliation{Instituut-Lorentz, Universiteit Leiden, P.O. Box 9506, 2300 RA Leiden, The Netherlands}
\date{April 2010}
\begin{abstract}
This is a numerical study of quasiparticle localization in symmetry class \textit{BD} (realized, for example, in chiral \textit{p}-wave superconductors), by means of a staggered-fermion lattice model for two-dimensional Dirac fermions with a random mass. For sufficiently weak disorder, the system size dependence of the average (thermal) conductivity $\sigma$ is well described by an effective mass $M_{\rm eff}$, dependent on the first two moments of the random mass $M(\bm{r})$. The effective mass vanishes linearly when the average mass $\bar{M}\rightarrow 0$, reproducing the known insulator-insulator phase boundary with a scale invariant dimensionless conductivity $\sigma_{c}=1/\pi$ and critical exponent $\nu=1$. For strong disorder a transition to a metallic phase appears, with larger $\sigma_{c}$ but the same $\nu$. The intersection of the metal-insulator and insulator-insulator phase boundaries is identified as a \textit{repulsive} tricritical point. 
\end{abstract}
\maketitle

\section{Introduction}
\label{intro}

\subsection{Description of the problem}
\label{description}

Superconductors with neither time-reversal symmetry nor spin-rotation symmetry (for example, having chiral \textit{p}-wave pairing) still retain one fundamental symmetry: the charge-conjugation (or particle-hole) symmetry of the quasiparticle excitations. Because of this symmetry, quasiparticle localization in a disordered chiral \textit{p}-wave superconductor is in a different universality class than in a normal metal.\cite{Eve08} The difference is particularly interesting in two dimensions, when the quantum Hall effect governs the transport properties. The electrical quantum Hall effect in a normal metal has the thermal quantum Hall effect as a superconducting analogue,\cite{Rea00,Sen00,Vis01} with different scaling properties because of the particle-hole symmetry.

The quasiparticles in a superconductor have electron and hole components $\psi_{e},\psi_{h}$ that are eigenstates, at excitation energy $\varepsilon$, of the Bogoliubov-De Gennes equation
\begin{equation}
\begin{pmatrix}
H_{0}-E_{F}&\Delta\\
\Delta^{\dagger}&-H^{\ast}_{0}+E_{F}
\end{pmatrix}\begin{pmatrix}
\psi_{e}\\ \psi_{h}
\end{pmatrix}=\varepsilon\begin{pmatrix}
\psi_{e}\\ \psi_{h}
\end{pmatrix}.\label{BdG}
\end{equation}
In a chiral \textit{p}-wave superconductor the order parameter $\Delta=\tfrac{1}{2}\{\chi(\bm{r}),p_{x}-ip_{y}\}$ depends linearly on the momentum $\bm{p}=-i\hbar\partial/\partial\bm{r}$, so the quadratic terms in the single-particle Hamiltonian $H_{0}=p^{2}/2m+U(\bm{r})$ may be neglected near $p=0$. For a uniform order parameter $\chi(\bm{r})=\chi_{0}$,  the quasiparticles are eigenstates of the Dirac Hamiltonian
\begin{equation}
H_{\rm Dirac}=v(p_{x}\sigma_{x}+p_{y}\sigma_{y})+v^{2}M(\bm{r})\sigma_{z},\label{HDirac}
\end{equation}
with velocity $v=\chi_{0}$ and mass $M=(U-E_{F})/\chi_{0}^{2}$ (distinct from the electron mass $m$). The particle-hole symmetry is expressed by
\begin{equation}
\sigma_{x}H_{\rm Dirac}^{\ast}\sigma_{x}=-H_{\rm Dirac}.\label{ehsymm}
\end{equation}
Randomness in the electrostatic potential $U(\bm{r})$ translates into randomness in the mass $M(\bm{r})=\bar{M}+\delta M(\bm{r})$ of the Dirac fermions. The sign of the average mass $\bar{M}$ determines the thermal Hall conductance,\cite{Rea00,Sen00,Vis01} which is zero for $\bar{M}>0$ (strong pairing regime) and quantized at $G_{0}=\pi^{2}k_{B}^{2}T/6h$ for $\bar{M}<0$ (weak pairing regime).

The thermal quantum Hall transition at $\bar{M}=0$ is analogous to the electrical quantum Hall transition at the center of a Landau level, but the scaling of the thermal conductivity $\sigma$ near the phase boundary is different from the scaling of the electrical conductivity because of the particle-hole symmetry. A further difference between these two problems appear if the superconducting order parameter contains vortices.\cite{Rea00,Boc00,Rea01} A vortex of unit vorticity contains a Majorana bound state at $\varepsilon=0$, in the weak-pairing regime.\cite{Vol99,Gur07} A sufficiently large density of Majorana bound states allows for extended states at the Fermi level ($\varepsilon=0$), with a thermal conductivity increasing $\propto\ln L$ with increasing system size $L$.\cite{Sen00} This socalled thermal metal has no counterpart in the electronic quantum Hall effect.

\begin{figure}[tb]
\centerline{\includegraphics[width=0.8\linewidth]{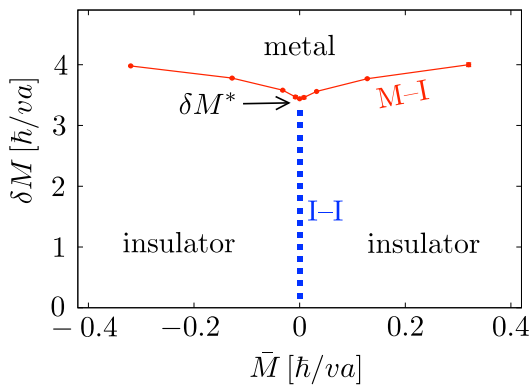}}
\caption{\label{fig_phasediagram}
Phase diagram in symmetry class \textit{BD}, calculated numerically from the lattice model of staggered fermions described in Sec.\ \ref{staggered}. (A qualitatively similar phase diagram was calculated for a different model\cite{Cho97} in Refs.\ \onlinecite{Cha02} and \onlinecite{Kag08}.) The thermal conductivity decays exponentially $\propto e^{-L/\xi}$ in the localized phase and increases $\propto\ln L$ in the metallic phase. The thermal conductivity is scale invariant on the metal-insulator (M--I) phase boundary (red solid line), as well as on the insulator-insulator (I--I) phase boundary (blue dashed line). The M--I and I--I phase boundaries meet at the tricritical point $\delta M^{\ast}$.
}
\end{figure}

As indicated in Fig.\ \ref{fig_phasediagram}, there are two types of phase transitions,\cite{Cha02,Kag08} a metal-insulator (M--I) transition upon decreasing $\delta M$ at constant $\bar{M}$ and an insulator-insulator (I--I) transition upon decreasing $\bar{M}$ through zero at constant (not too large) $\delta M$. The I--I transition separates phases with a different value of the thermal Hall conductance, while the M--I transition separates the thermal metal from the thermal insulator. Only the I--I transition remains if there are no vortices, or more generally, if there are no Majorana bound states.\cite{Rea00,Boc00,Rea01} In the nomenclature of Ref.\ \onlinecite{Boc00}, the symmetry class is called \textit{BD} with Majorana bound states and \textit{D} without.

\subsection{Objectives of this paper}
\label{objectives}

The primary purpose of our paper is to investigate, by numerical simulation, to what extent the scale dependence of localization by a random mass can be described in terms of an effective non-fluctuating mass: $\sigma(L,\bar{M},\delta M)=\sigma(L,M_{\rm eff},0)$, for some function $M_{\rm eff}(\bar{M},\delta M)$. Because there is no other length scale in the problem at zero energy, $\sigma(L,M_{\rm eff},0)$ can only depend on $L$ and $M_{\rm eff}$ through the dimensionless combination $LM_{\rm eff}v/\hbar\equiv L/\xi$. The effective-mass hypothesis thus implies one-parameter scaling: $\sigma(L,\bar{M},\delta M)=\sigma_{0}(L/\xi)$. Two further implications concern the critical conductivity $\sigma_{c}$ (which is the scale invariant value of $\sigma$ on the phase boundary $\bar{M}=0$) and the critical exponent $\nu$ (governing the divergence of the localization length $\xi\propto\bar{M}^{-\nu}$). 

Both $\sigma_{c}$ and $\nu$ follow directly from the effective mass hypothesis. By construction, the scaling function $\sigma_{0}$ is the conductivity of ballistic massless Dirac fermions, which has been calculated in the context of graphene. For a system with dimensions $L\times W$, and periodic boundary conditions in the transverse direction, it is given by\cite{Kat06,Two06}
\begin{align}
&\sigma_{0}(L/\xi)=G_{0}\frac{L}{W}\sum_{n=-\infty}^{\infty}\cosh^{-2}\sqrt{(2\pi nL/W)^2+(L/\xi)^{2}}\nonumber\\
&\qquad\xrightarrow{W\gg L}G_{0}\frac{1}{\pi}\int_{0}^{\infty}dq\,\cosh^{-2}\sqrt{q^2+(L/\xi)^{2}}.\label{sigma0result}
\end{align}
A scale invariant conductivity
\begin{equation}
\lim_{\xi\rightarrow\infty}\sigma_{0}(L/\xi)\equiv\sigma_{c}=G_{0}\frac{L}{W}\sum_{n=-\infty}^{\infty}\cosh^{-2}(2\pi nL/W)\label{sigmacsum}
\end{equation}
is reached for vanishing effective mass. In the limit of a large aspect ratio $W/L\gg 1$ we recover the known value $\sigma_{c}=G_{0}/\pi$ of the critical conductivity for a random mass with zero average.\cite{Lud94} The critical exponent $\nu=1$ follows by comparing the expansion of the conductivity
\begin{equation}
\sigma(L,\bar{M},\delta M)=\sigma_{c}+[L^{1/\nu}\bar{M}f(\delta M)]^{2}+{\cal O}(\bar{M})^{4}\label{sigmaexpansion}
\end{equation}
in (even) powers of $\bar{M}$ with the expansion of the scaling function \eqref{sigma0result} in powers of $L$. This value for $\nu$ is aspect-ratio independent and agrees with the known result for the I--I transition.\cite{Eve08}

The description in terms of an effective mass breaks down for strong disorder. We find that the scaling function at the M--I transition differs appreciably from $\sigma_{0}$, with an aspect-ratio independent critical conductivity $\sigma_{c}\approx 0.4\,G_{0}$. The critical exponent remains close to or equal to $\nu=1$ (in disagreement with earlier numerical simulations \cite{Kag08}).

The secondary purpose of our paper is to establish the nature of the tricritical point $\delta M^{\ast}$ at which the two insulating phases and the metallic phase meet. From the scale dependence of $\sigma$ near this tricritical point, we conclude that it is a \textit{repulsive} fixed point (in the sense that $\sigma$ scales with increasing $L$ to larger values for $\delta M>\delta M^{\ast}$ and to smaller values for $\delta M<\delta M^{\ast}$). An \textit{attractive} tricritical point had been suggested as a possible scenario,\cite{Mil07,Kag09} in combination with a repulsive critical point at some $\delta M^{\ast\ast}<\delta M^{\ast}$. Our numerics does not support this scenario.  

\subsection{Outline}
\label{outline}

The outline of this paper is as follows. In the next Section we summarize the lattice fermion model that we use to simulate quasiparticle localization in symmetry class \textit{BD}. We only give a brief description, referring to Refs.\ \onlinecite{Two08} and \onlinecite{Wim10} for a more detailed presentation of the model. The scaling of the thermal conductivity and the localization length near the insulator-insulator and metal-insulator transitions are considered separately in Secs.\ \ref{IItransition} and \ref{MItransition}, respectively. The tricritical point, at which the two phase boundaries meet, is studied in Sec.\ \ref{tricritical}. We conclude in Sec.\ \ref{conclude}.

\section{Staggered fermion model}
\label{staggered}

Earlier numerical investigations\cite{Cha02,Mil07,Kag08,Kag09} of the class \textit{BD} phase diagram were based on the Cho-Fisher network model.\cite{Cho97} Here we use a staggered fermion model in the same symmetry class, originally developed in the context of lattice gauge theory\cite{Sta83,Ben83} and recently adapted to the study of transport properties in graphene.\cite{Two08}  An attractive feature of the lattice model is that, by construction, it reduces to the Dirac Hamiltonian on length scales large compared to the lattice constant $a$.

The model is defined on a square lattice in a strip geometry, extending in the longitudinal direction from $x=0$ to $x=L=N_{x}a$ and in the transverse direction from $y=0$ to $y=W=N_{y}a$. We use  periodic boundary conditions in the transverse direction. The transfer matrix ${\cal T}$ from $x=0$ to $x=L$  is derived in Ref.\ \onlinecite{Two08}, and we refer to that paper for explicit formulas. The dispersion relation of the staggered fermions,
\begin{equation}
\tan^{2}(k_{x}a/2)+\tan^{2}(k_{y}a/2)+\left(\frac{Mav}{2\hbar}\right)^2=\left(\frac{\varepsilon a}{2\hbar v}\right)^{2},\label{dispersion}
\end{equation}
has a Dirac cone at wave vectors $|\bm{k}|a\ll 1$ which is gapped by a nonzero mass. Staggered fermions differ from Dirac fermions by the pole at the edge of Brillouin zone ($|k_{x}|\rightarrow\pi/a$ or $|k_{y}|\rightarrow\pi/a$), which is insensitive to the presence of a mass. We do not expect these large-wave number modes to affect the large-length scaling of the conductivity, because they preserve the electron-hole symmetry. 

The energy is fixed at $\varepsilon=0$ (corresponding to the Fermi level for the superconducting quasiparticles). The transfer matrix ${\cal T}$ is calculated recursively using a stable QR decomposition algorithm.\cite{Kra05} An alternative stabilization method\cite{Two08} is used to recursively calculate the transmission matrix $t$. Both algorithms give consistent results, but the calculation of ${\cal T}$ is more accurate than that of $t$ because it preserves the electron-hole symmetry irrespective of round-off errors.

The random mass is introduced by randomly choosing values of $M$ on each site uniformly in the interval $(\bar M-\delta M,\bar M+\delta M)$. Variations of $M(\bm{r})$ on the scale of the lattice constant introduce Majorana bound states, which place the model in the \textit{BD} symmetry class.\cite{Wim10} In principle, it is possible to study also the class \textit{D} phase diagram (without Majorana bound states), by choosing a random mass landscape that is smooth on the scale of $a$. Such a study was recently performed,\cite{Bar10} using a different model,\cite{Bar07} to demonstrate the absence of the M--I transition in class \textit{D}.\cite{Rea00,Boc00,Rea01} Since here we wish to study both the I--I and M--I transitions, we do not take a smooth mass landscape.

\section{Scaling near the insulator-insulator transition}
\label{IItransition}

\subsection{Scaling of the conductivity}
\label{sigmaIIscaling}

\begin{figure}[tb]
\centerline{\includegraphics[width=0.8\linewidth]{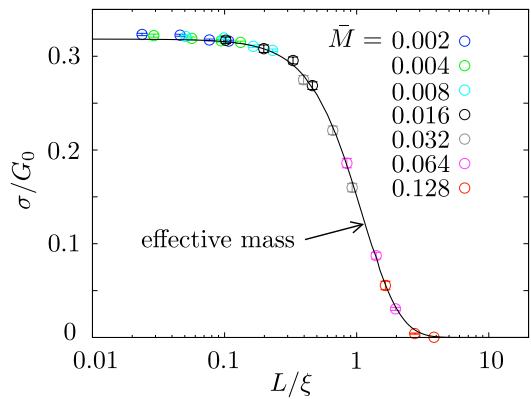}}
\caption{\label{fig_sigmaII}
Average conductivity $\sigma$ (with error bars indicating the statistical uncertainty) at fixed disorder strength $\delta M=2.5\,\hbar /va$, as a function of system size $L$. The aspect ratio of the disordered strip is fixed at $W/L=5$. Data sets at different values of $\bar{M}$ (listed in the figure in units of $\hbar/va$) collapse upon rescaling by $\xi$ onto a single curve (solid line), given by Eq.\ \eqref{sigma0result} in terms of an effective mass $M_{\rm eff}=\hbar/v\xi$.
}
\end{figure}

\begin{figure}[tb]
\centerline{\includegraphics[width=0.8\linewidth]{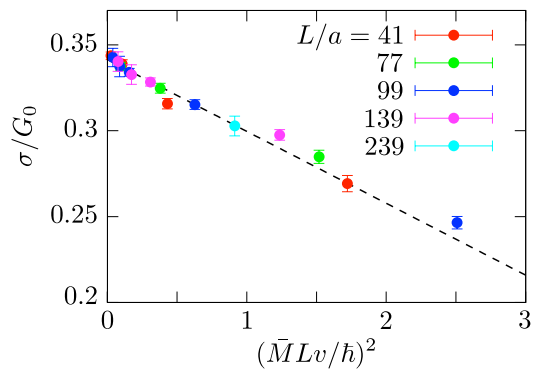}}
\caption{\label{fig_scalingII}
Plot of the average conductivity $\sigma$ versus $(\bar{M}L)^{2}$, for fixed $\delta M=2.5\,\hbar/va$ and $W/L=3$. The dashed line is a least-square fit through the data, consistent with critical exponent $\nu=1$.
}
\end{figure}

In Fig.\ \ref{fig_sigmaII} we show the average (thermal) conductivity $\sigma=(L/W)\langle{\rm Tr}\,tt^{\dagger}\rangle$ (averaged over some $10^{3}$ disorder realizations) as a function of $L$ for a fixed $\delta M$ in the localized phase. Data sets with different $\bar{M}$ collapse on a single curve upon rescaling with $\xi$. (In the logarithmic plot this rescaling amounts simply to a horizontal displacement of the entire data set.) The scaling curve (solid line in Fig.\ \ref{fig_sigmaII}) is the effective mass conductivity \eqref{sigma0result}, with $M_{\rm eff}=\hbar/v\xi$. Fig.\ \ref{fig_scalingII} shows the linear scaling of $\sigma$ with $(\bar{M}L)^{2}$ for small $\bar{M}$, as expected from Eq.\ \eqref{sigmaexpansion} with $\nu=1$.

\begin{figure}[tb]
\centerline{\includegraphics[width=0.8\linewidth]{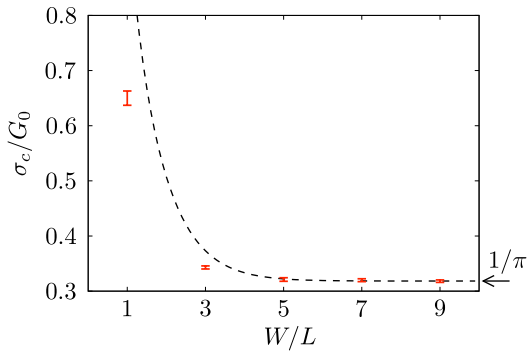}}
\caption{\label{fig_sigmaaspect}
Dependence on the aspect ratio $W/L$ of the critical conductivity at the insulator-insulator transition ($\bar{M}=0$). The disorder strength is fixed at $\delta M=2.5\,\hbar /va$. The dashed curve is the aspect ratio dependence of the conductivity of ballistic massless Dirac fermions [Eq.\ \eqref{sigmacsum}].
}
\end{figure}

We have studied the aspect ratio dependence of the critical conductivity $\sigma_{c}$. As illustrated in Fig.\ \ref{fig_sigmaaspect}, the convergence for $W/L\rightarrow\infty$ is to the value $\sigma_{c}=1/\pi$ expected from Eq.\ \eqref{sigma0result}. The conductivity of ballistic massless Dirac fermions also has an aspect ratio dependence,\cite{Two06} given by Eq.\ \eqref{sigmacsum} (for periodic boundary conditions). The comparison in Fig.\ \ref{fig_sigmaaspect} of $\sigma_{c}$ with Eq.\ \eqref{sigmacsum} shows that $\sigma_{c}$ has a somewhat weaker aspect ratio dependence. 

\subsection{Scaling of the Lyapunov exponent}
\label{LyapunovIIscaling}

The transfer matrix ${\cal T}$ provides an independent probe of the critical scaling through the Lyapunov exponents. The transfer matrix product ${\cal TT}^{\dagger}$ has eigenvalues $e^{\pm\mu_{n}}$ with $0\leq\mu_{1}\leq\mu_{2}\leq\cdots$. The $n$-th Lyapunov exponent $\alpha_{n}$ is defined by
\begin{equation}
\alpha_{n}=\lim_{L\rightarrow\infty}\frac{\mu_{n}}{L}.\label{alphadef}
\end{equation}
The dimensionless product $W\alpha_{1}\equiv\Lambda$ is the inverse of the MacKinnon-Kramer parameter.\cite{Mac81} We obtain $\alpha_{1}$ by increasing $L$ at constant $W$ until convergence is reached (typically for $L/W\simeq 10^{3}$). The large-$L$ limit is self-averaging, but some improvement in statistical accuracy is reached by averaging over a small number (10--20) of disorder realizations.

\begin{figure}[tb]
\centerline{\includegraphics[width=0.8\linewidth]{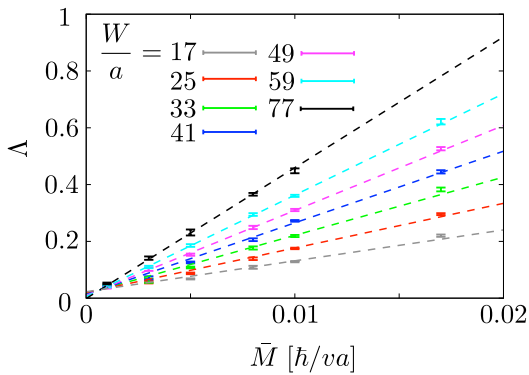}}
\caption{\label{fig_lambdascalingII}
Plot of $\Lambda=W\alpha_{1}$ (with $\alpha_{1}$ the first Lyapunov exponent) as a function of $\bar{M}$ near the insulator-insulator transition, for fixed $\delta M=2.5\,\hbar v/a$ and different values of $W$. The dashed lines are a fit to Eq.\ \eqref{Lambdascaling}.
}
\end{figure}

We seek the coefficients in the scaling expansion
\begin{equation}
\Lambda=\Lambda_{c}+c_{1}W^{1/\nu}(\bar{M}-M_{c})+{\cal O}(\bar{M}-M_{c})^{2},\label{Lambdascaling}
\end{equation}
for fixed $\delta M$. The fit in Fig.\ \ref{fig_lambdascalingII} gives $\Lambda_{c}=0.03$, $\nu=1.05$, $M_{c}=7\cdot 10^{-4}$, consistent with the expected values\cite{Cha02} $\Lambda_{c}=0$, $\nu=1$, $M_{c}=0$.

\section{Scaling near the metal-insulator transition}
\label{MItransition}

\subsection{Scaling of the conductivity}
\label{sigmaMIscaling}

\begin{figure}[tb]
\centerline{\includegraphics[width=0.9\linewidth]{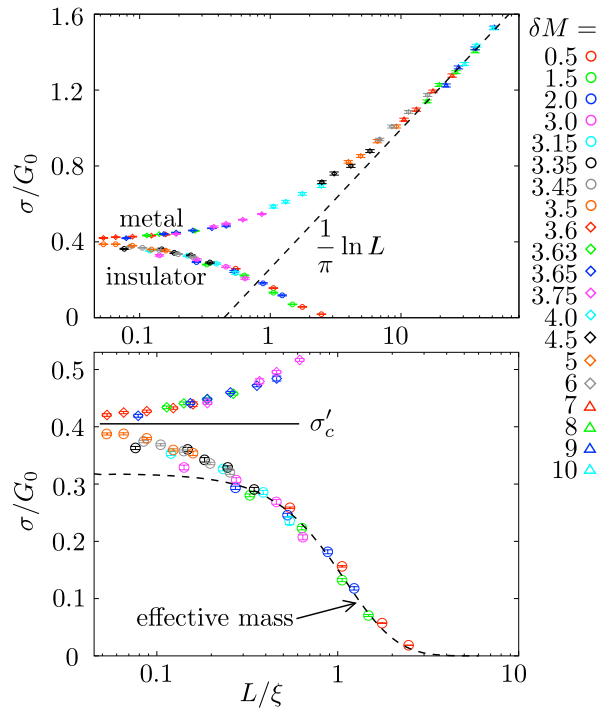}}
\caption{\label{fig_scalingMI}
Average conductivity $\sigma$ at fixed average mass $\bar{M}=0.032\,\hbar /va$, as a function of system size $L$. (The two panels show the same data on a different scale.) The aspect ratio of the disordered strip is fixed at $W/L=5$. Data sets at different values of $\delta M$ (listed in the figure in units of $\hbar/va$) collapse upon rescaling by $\xi$ onto a pair of curves in the metallic and insulating regimes. The metal-insulator transition has a scale invariant conductivity $\sigma'_{c}$, larger than the value $G_{0}/\pi$ which follows from the effective mass scaling (dashed curve in the lower panel). The upper panel shows that the conductivity in the metallic regime follows the logarithmic scaling \eqref{sigmalnL}.
}
\end{figure}

To investigate the scaling near the metal-insulator transition, we increase $\delta M$ at constant $\bar{M}$. Results for the conductivity are shown in Fig.\ \ref{fig_scalingMI}. In the metallic regime $\delta M>\delta M_{c}$ the conductivity increases logarithmically with system size $L$, in accord with the theoretical prediction:\cite{Sen00,Eve08}
\begin{equation}
\sigma/G_{0}=\frac{1}{\pi}\ln L+{\rm constant}.\label{sigmalnL}
\end{equation}
(See the dashed line in Fig.\ \ref{fig_scalingMI}, upper panel.)

In the insulating regime $\delta M<\delta M_{c}$ the conductivity decays exponentially with system size, while it is scale independent at the critical point $\delta M=\delta M_{c}$. Data sets for different $\delta M$ collapse onto a single function of $L/\xi$, but this function is different from the effective mass scaling $\sigma_{0}(L/\xi)$ of Eq.\ \eqref{sigma0result}. (See the dashed curve in Fig.\ \ref{fig_scalingMI}, lower panel.) This indicates that the effective mass description, which applies well near the insulator-insulator transition, breaks down at large disorder strengths near the metal-insulator transition. The two transitions therefore have a different scaling behavior, and can have different values of critical conductivity and critical exponent (which we denote by $\sigma'_{c}$ and $\nu'$).

\begin{figure}[tb]
\centerline{\includegraphics[width=0.8\linewidth]{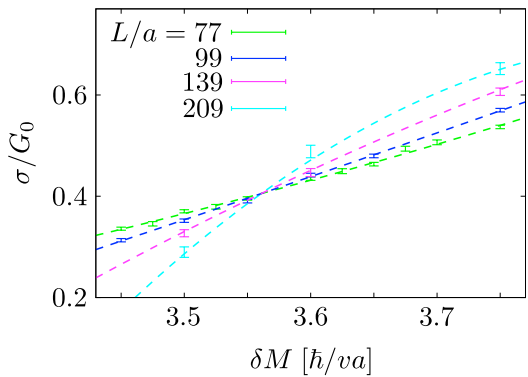}}
\caption{\label{fig_sigmaMI}
Plot of the average conductivity $\sigma$ as a function of $\delta M$ near the metal-insulator transition, for fixed $\bar{M}=0.032\,\hbar/va$. The length $L$ is varied at fixed aspect ratio $W/L=3$. The dashed curves are a fit to Eq.\ \eqref{sigmafit2}.
}
\end{figure}

Indeed, the critical conductivity $\sigma'_{c}=0.41\,G_{0}$ is significantly larger than the ballistic value $G_{0}/\pi=0.32\,G_{0}$. Unlike at the insulator-insulator transition, we found no significant aspect-ratio dependence in the value of $\sigma'_{c}$. To obtain the critical exponent $\nu'$ we follow Ref.\ \onlinecite{Asa04} and fit the conductivity near the critical point including terms of second order in $\delta M-\delta M_{c}$:
\begin{align}
\sigma={}&\sigma'_{c}+c_{1}L^{1/\nu'}[\delta M-\delta M_{c}+c_{2}(\delta M-\delta M_{c})^{2}]\nonumber\\
&+c_{3}L^{2/\nu'}(\delta M-\delta M_{c})^{2}.\label{sigmafit2}
\end{align}
Results are shown in Fig.\ \ref{fig_sigmaMI}, with $\nu'=1.02\pm 0.06$. The quality of the multi-parameter fit is assured by a reduced chi-squared value close to unity ($\chi^{2}=0.95$). Within error bars, this value of the critical exponent is the same as the value $\nu=1$ for the insulator-insulator transition.

\subsection{Scaling of the Lyapunov exponent}
\label{LyapunovMIscaling}

\begin{figure}[tb]
\centerline{\includegraphics[width=0.8\linewidth]{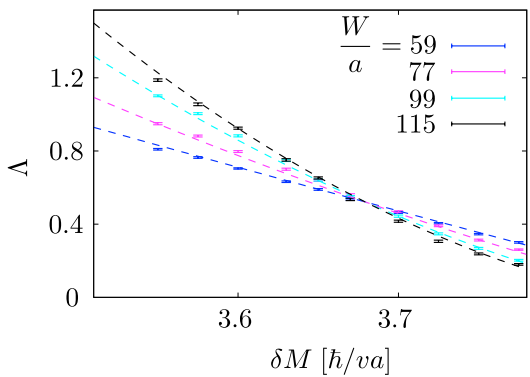}}
\caption{\label{fig_lambdaMI}
Plot of $\Lambda=W\alpha_{1}$ (with $\alpha_{1}$ the first Lyapunov exponent) as a function of $\delta M$ near the metal-insulator transition, for fixed $\bar{M}=0.032\,\hbar v/a$ and different values of $W$. The dashed curves are a fit to Eq.\ \eqref{lambdafit2}.
}
\end{figure}

As an independent measurement of $\nu'$, we have investigated the finite-size scaling of the first Lyapunov exponent. Results are shown in Fig.\ \ref{fig_lambdaMI}. Within the framework of single-parameter scaling, the value of $\nu'$ should be the same for $\sigma$ and $\Lambda$, but the other coefficients in the scaling law may differ,
\begin{align}
\Lambda={}&\Lambda_{c}+c'_{1}L^{1/\nu'}[\delta M-\delta M'_{c}+c'_{2}(\delta M-\delta M'_{c})^{2}]\nonumber\\
&+c'_{3}L^{2/\nu'}(\delta M-\delta M'_{c})^{2}.\label{lambdafit2}
\end{align}
Results are shown in Fig.\ \ref{fig_lambdaMI}, with $\nu'=1.06\pm0.05$. The chi-squared value for this fit is relatively large, $\chi^{2}=5.0$, but the value of $\nu'$ is consistent with that obtained from the conductivity (Fig.\ \ref{fig_sigmaMI}).

\section{Tricritical point}
\label{tricritical}

\begin{figure}[tb]
\centerline{\includegraphics[width=0.8\linewidth]{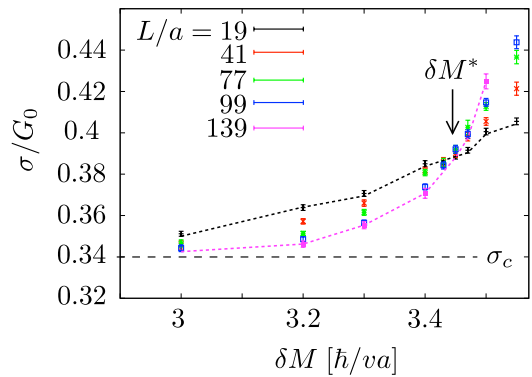}}
\caption{\label{fig_tricritical}
Conductivity $\sigma$ as a function of $\delta M$ on the critical line $\bar{M}=0$, for different values of $L$ at fixed aspect ratio $W/L=3$. (The dotted lines through data points are guides to the eye.) The tricritical point $\delta M^{\ast}$ is indicated, as well as the scale invariant large-$L$ limit $\sigma_{c}$ for $\delta M<\delta M^{\ast}$.
}
\end{figure}

As indicated in the phase diagram of Fig.\ \ref{fig_phasediagram}, the tricritical point at $\bar{M}=0$, $\delta M=\delta M^{\ast}$ is the point at which the insulating phases at the two sides of the I--I transition meet the metallic phase. We have searched for this tricritical point by calculating the scale dependence of the conductivity $\sigma$ on the line $\bar{M}=0$ for different $\delta M$. Results are shown in Fig.\ \ref{fig_tricritical}.

The calculated scale dependence is consistent with the identification of the point $\delta M^{\ast}=3.44\,\hbar/va$ as a \textit{repulsive} fixed point. The conductivity increases with increasing $L$ for $\delta M>\delta M^{\ast}$, while for $\delta M<\delta M^{\ast}$ it decreases towards the scale invariant large-$L$ limit $\sigma_{c}$.

\section{Discussion}
\label{conclude}

We have studied quasiparticle localization in symmetry class \textit{BD}, by means of a lattice fermion model.\cite{Two08} The thermal quantum Hall effect\cite{Rea00,Sen00,Vis01} in a chiral \textit{p}-wave superconductor at weak disorder is in this universality class, as is the phase transition to a thermal metal\cite{Sen00} at strong disorder. For weak disorder our lattice model can also be used to describe the localization of Dirac fermions in graphene with a random gap\cite{Bar10,Zie09,Zie10} (with $\sigma$ the electrical, rather than thermal, conductivity and $G_{0}=4e^{2}/h$ the electrical conductance quantum). The metallic phase at strong disorder requires Majorana bound states,\cite{Rea00,Boc00,Rea01} which do not exist in graphene (symmetry class \textit{D} rather than \textit{BD}). We therefore expect the scaling analysis in Sec.\ \ref{IItransition} at the insulator-insulator (I--I) transition to be applicable to chiral \textit{p}-wave superconductors as well as to graphene, while the scaling analysis of Sec.\ \ref{MItransition} at the metal-insulator (M--I) transition applies only in the context of superconductivity.

Our lattice fermion model is different from the network model\cite{Cho97} used in previous investigations,\cite{Cha02,Mil07,Kag08,Kag09} but it falls in the same universality class so we expect the same critical conductivity and critical exponent. For the I--I transition analytical calculations\cite{Eve08,Lud94} give $\sigma_{c}=G_{0}/\pi$ and $\nu=1$, in agreement with our numerics. There are no analytical results for the M--I transition. We find a slightly larger critical conductivity ($\sigma'_{c}=0.4\,G_{0}$), which has the qualitatively more significant consequence that the effective mass scaling which we have demonstrated at the I--I transition breaks down at the M--I transition (compare Figs.\ \ref{fig_sigmaII} and \ref{fig_scalingMI}, lower panel). 

We conclude from our numerics that the critical exponents $\nu$ at the I--I transition and $\nu'$ at the M--I transition are both equal to unity within a 5\% error margin, which is significantly smaller than the result $\nu=\nu'=1.4\pm 0.2$ of an earlier numerical investigation.\cite{Kag08} It is much more difficult to calculate accurately the critical exponent than the conductivity, so we do not attach too much significance to this disagreement. (Athough we find it puzzling that the value for $\nu$ obtained in Ref.\ \onlinecite{Kag08} lies so much higher than the analytical prediction $\nu=1$.) The logarithmic scaling \eqref{sigmalnL} of the conductivity in the thermal metal phase, predicted analytically,\cite{Sen00,Eve08} is nicely reproduced by our numerics (Fig.\ \ref{fig_scalingMI}, upper panel). 

The nature of the tricritical point has been much debated in the literature.\cite{Mil07,Kag09} Our numerics indicates that this is a repulsive critical point (Fig.\ \ref{fig_tricritical}). This finding lends support to the simplest scaling flow along the I--I phase boundary,\cite{Lud94} towards the free-fermion fixed point at $\bar{M}=0$, $\delta M=0$.

In conclusion, we hope that this investigation brings us further to a complete understanding of the phase diagram and scaling properties of the thermal quantum Hall effect. We now have two efficient numerical models in the \textit{BD} universality class, the Cho-Fisher network model\cite{Cho97} studied previously and the lattice fermion model\cite{Two08} studied here. There is a consensus on the scaling at weak disorder, although some disagreement on the scaling at strong disorder remains to be resolved.

\acknowledgments

We have benefited from discussions with A. R. Akhmerov, J. H. Bardarson, C. W. Groth, and M. Wimmer. This research was supported by the Dutch Science Foundation NWO/FOM, by an ERC Advanced Investigator Grant, by the EU network NanoCTM, and by the ESF network EuroGraphene.

\end{document}